\pdfoutput=1
\documentclass[amsymb,twocolumn,prb,showpacs,aps,superscriptaddress]{revtex4-1}

\usepackage[english]{babel}
\usepackage[ansinew]{inputenc}
\usepackage{times}
\usepackage{float}
\usepackage{graphicx}
\usepackage{graphics}
\usepackage{amsmath}
\usepackage{amsfonts}
\usepackage{amssymb}
\usepackage{amsbsy}
\usepackage{epstopdf}
\usepackage{makeidx}
\usepackage[caption=false]{subfig}
\usepackage{ytableau}
\usepackage[export]{adjustbox}
\usepackage{algpseudocode}
\usepackage{color}
\usepackage{pgf}
\usepackage{bm}
\usepackage{longtable}
\usepackage{appendix}
\usepackage{tikz} 
\usepackage[normalem]{ulem}
\usepackage{multirow,bigstrut}

\newcommand{\be}{\begin{equation}}
\newcommand{\ee}{\end{equation}}
\newcommand{\bea}{\begin{eqnarray}}
\newcommand{\eea}{\end{eqnarray}}

\usepackage{color,soul}

\makeindex

\begin{document}

\title{Generalized representative structures for atomistic systems}

\author{James M. Goff}
\affiliation{Center for Computing Research, Sandia National Laboratories, Albuquerque, New Mexico 87185, USA}
\author{Coreen Mullen}
\affiliation{Center for Computing Research, Sandia National Laboratories, Albuquerque, New Mexico 87185, USA}
\affiliation{Department of Mathematics, Computer Science, and Engineering Technology, Elizabeth City State University, Elizabeth City, North Carolina 27909, USA}
\author{Shizhong Yang}
\affiliation{Computer Science Department, Southern University, Baton Rouge, Louisiana 70807, USA}
\author{Oleg N. Starovoytov}
\affiliation{Computer Science Department, Southern University, Baton Rouge, Louisiana 70807, USA}
\affiliation{Center for Computation and Technology, Louisiana State University, Baton Rouge, Louisiana 70803, USA}

\author{Mitchell A. Wood}
\affiliation{Center for Computing Research, Sandia National Laboratories, Albuquerque, New Mexico 87185, USA}
%


\begin{abstract}
A new method is presented to generate atomic structures that reproduce the essential characteristics of arbitrary material systems, phases, or ensembles. Previous methods allow one to reproduce the essential characteristics (e.g., the chemical disorder) of a large random alloy within a small crystal structure. 
The ability to generate small representations of random alloys, along with the restriction to crystal systems, results from using the fixed-lattice cluster correlations to describe structural characteristics. A more general description of the structural characteristics of atomic systems is obtained using complete sets of atomic environment descriptors.
These are used within for generating representative atomic structures without restriction to fixed lattices.
A general data-driven approach is provided here utilizing the atomic cluster expansion (ACE) basis. The $N$-body ACE descriptors are a complete set of atomic environment descriptors that span both chemical and spatial degrees of freedom and are used within for describing atomic structures.
The generalized representative structure (GRS) method presented within generates small atomic structures that reproduce ACE descriptor distributions corresponding to arbitrary structural and chemical complexity. 
It is shown that systematically improvable representations of crystalline systems on fixed parent lattices, amorphous materials, liquids, and ensembles of atomic structures may be produced efficiently through optimization algorithms. 
With the GRS method, we highlight reduced representations of atomistic machine-learning training datasets that contain similar amounts of information and small 40-72 atom representations of liquid phases. The ability to use GRS methodology as a driver for informed novel structure generation is also demonstrated. The advantages over other data-driven methods and state-of-the-art methods restricted to high-symmetry systems are highlighted.

\end{abstract}


\maketitle

\section{\label{intro}Introduction}
Curation of databases worth of physical, chemical, and material property data has accelerated the progress in many areas of science and has grown in importance simply by avoiding duplication of effort.\cite{jain2013commentary,talirz2020materials} 
This is particularly true for computational methods where accurate prediction of even simple properties can require large resources, or where candidate materials cannot be (or have not yet been) synthesized in the laboratory. 
Often these properties are predicted for perfect crystals or simple molecules due to the ability to enumerate large numbers of candidates. \cite{pence2010chemspider,kim2016pubchem}
These databases do not fully enumerate every possible structural and chemical combination that is possible; this is particularly true for chemically complex systems and materials that do not exhibit crystalline structure. 
A secondary outcome of these computational databases is the development of efficient surrogate models that can reduce the computational cost of subsequent property predictions.
The entries in the databases limited to the lowest energy and highest symmetry atomic structures is not sufficient for many surrogate property models.

The type of data in these databases such as those in materials project\cite{jain2013commentary} and materials cloud\cite{talirz2020materials} varies significantly based on the end application. 
These databases contain high-fidelity electronic structure data and thermodynamic properties associated with crystalline atomic structures. Databases comprised of simple crystalline atomic structures (structures with long-range symmetry belonging to specific space groups) are often made through combinatorial searches of crystal structures over different space groups.
Other datasets and databases are tailored to train larger-scale models. These include machine-learned interatomic potentials, thermodynamic energy models for atomistic systems, and other more data-driven studies.\cite{otis2017pycalphad,drautz_atomic_2019,goff2021predicting,xiong2021data} 


One highly successful family of materials models result directly from first-principles electronic structure theory. 
Typically density functional theory (DFT) is restricted to small-scale $~ 10^0 -10^2$-atom systems due to its expensive (often cubic) scaling with the number of electrons in the system. 
This often limits its direct application for integrated properties measured over time, favoring single structure measurements. 
Regardless, DFT is an extremely valuable tool to predict important properties in small systems.
Efforts to represent realistic materials and larger-scale phenomena in DFT length scales are often restricted to fixed lattices.\cite{puchala2023casm,zunger1990special,morgan2017generating,sorkin2020generalized} Thus these state-of-the-art methods, are not well-suited for liquid, amorphous, organics, and many other systems. 
In general, electronic structure methods are also prohibitively expensive for materials with large compositional degrees of freedom (such as high-entropy alloys).
A number of atomic structure generation methods that rely on \textit{a priori} energy models may also sample low-symmetry and compositionally complex materials, including some off-lattice structures.\cite{lyakhov2013new,larsen2017atomic} Such methods are limited by the energy model chosen, and in cases where DFT is the underlying energy model, can be computationally expensive.

Data-driven structure generation methods are compelling alternatives and generally do not require an \textit{a priori} energy model. Data-driven methods are therefore not limited to sampling states from a potential energy surface or bound by the expense of a high-fidelity energy model.
Minimum energy searches on a known potential energy surface will generate structures for local minima, but often times the transition states between minima are equally important, and will be omitted outright. 
Searching for structures based on symmetry arguments\cite{zhao2023physics} are robust, but are still limited to more ordered structures. 
Some data-driven methods are focused on providing diverse atomic information, such as the entropy maximization method.\cite{karabin2020entropy} These methods uniformly span a large volume of descriptor space but achieve this by producing hundreds of thousands of atomic structures.
Generating database properties (DFT calculations) for these large data sets is cumbersome, and some of the information superfluous in many applications.
In fact, when these structure generation tools are used for machine-learned potentials, it is often still combined with expert curated structures that are closer to ground state atomic configurations. 
Beyond possibly oversampling unrealistic structures, the magnitude of atomic structures produced with this method is not suitable for many computational budgets. 

\begin{figure}
    \centering
    \includegraphics[width=0.94\linewidth]{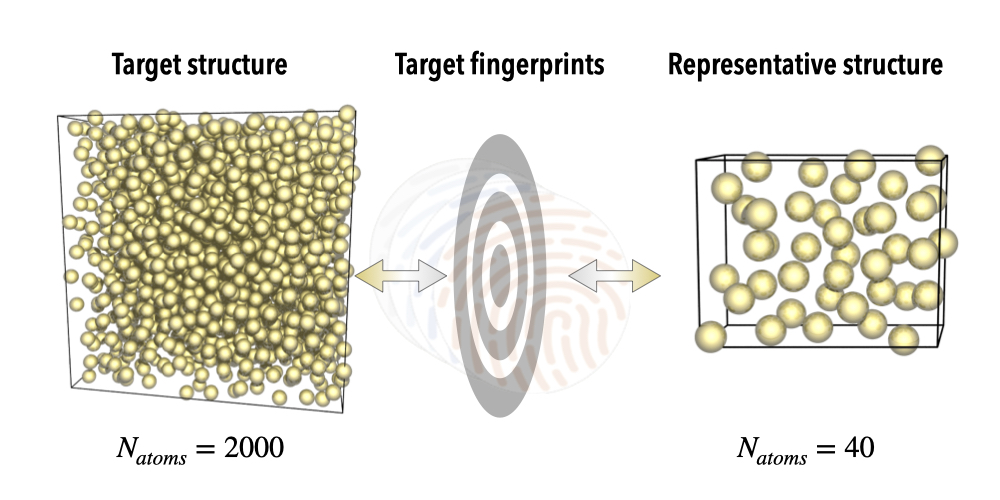}
    \caption{Illustration of the generalized representative structure method where a complex target distribution of atomic environment descriptors (a.k.a. fingerprints) may be set and a small structure that best reproduces those fingerprints is generated.}
    \label{fig:grs_ill}
\end{figure}

In this work, we present a new data-driven method for atomic structure generation, which is illustrated in Fig. \ref{fig:grs_ill}. It relies on the use of atomic cluster expansion (ACE) descriptors (a.k.a. atomic environment fingerprints) that are continuous and differentiable with respect to atomic positions.
These descriptors may be generated for arbitrary body order and form a complete basis for atomic environments with spatial and chemical degrees of freedom.\cite{drautz_atomic_2019,dusson_atomic_2022,goff2024permutation} 
The central idea behind the methods presented within is that structure(s) with specific ACE descriptor distributions may be generated without an \textit{a priori} energy model.
Here we demonstrate the generalized representative structure (GRS) method and how to generate arbitrary atomic structures characterized by the descriptor distributions with systematically improvable resolution and accuracy.
Data reduction demonstrations are given for single large structures as well as ensembles of structures.
Prior work attempted these outcomes with incomplete sets of descriptors (e.g., descriptors that are limited to low body order interactions).\cite{fung2022atomic} 
However, some structures and motifs are symmetrically indistinguishable from others when only using low-body order atomic environment descriptors, or incomplete basis sets.\cite{pozdnyakov2020incompleteness} 

The arbitrary body order of ACE and similar N-body descriptor sets may be used for complete representations of atomic structures.\cite{nigam2024completeness} This offers fundamental improvements over structure generation methods relying on incomplete descriptor sets.
ACE descriptors may be constructed such that they contain the same chemical information as fixed-lattice cluster functions, and the systematically improvable radial and angular resolution of N-body ACE descriptors also encodes arbitrarily complex information about variable atomic positions. 
We exploit these properties of ACE to develop the GRS method; target distributions of ACE descriptors may correspond to arbitrary atomic structures. 
Thus, the key advance provided by the GRS method is the ability to represent and generate atomic systems more broadly. In theory the completeness of the ACE descriptors allows one to do this universally for any atomic structure. 
However, we restrict ourselves to more simple systems for purposes of demonstration, and note that it may be extended to others.
For example we find structures that best represent alloys on a fixed parent lattice, liquids, amorphous systems, and ensembles of atomic structures.  
Where the goal is to predict properties from first-principles methods the GRS method is well suited to generate small atomic structures that represent large-scale atomistic systems and ensembles more generally than state-of-the-art methods for fixed lattices.
Like previous state-of-the-art methods for fixed lattices, this may be done with systematically improvable accuracy in how well the structures represent the target.


The GRS method is presented and defined within, including connections to relevant theory for fixed-lattice and atomic cluster expansions\cite{zunger1990special,drautz_atomic_2019}. 
GRS structure search is an evolutionary process where atoms experience forces that steer them toward the target structure, therefore this dynamic process is subject to convergence challenges.
Testing of GRS includes validation against random walks for examples where the exact solutions are known.
Starting with structurally complex representations, the first key application of GRS provides small, DFT-compatible representative atomic structures that represent a liquid Ta phase.
Exploring chemical complexity, the GRS method is used to generate new atomic structures to make binary alloy (Fe-Cr) datasets that are more compatible with phase changes and melting. 
Directionality of the search is then reversed where structures are generated such that they are unlike the descriptor distribution of a random $Fe_{0.35}Cr_{0.35}Si_{0.15}V_{0.15}$ alloy. 
This presents a systematic way to generate training data in a large composition/position space where the amount of new atomistic/chemical information can be tuned.
It is also shown that the GRS method is particularly useful for systematically extracting information from large-scale simulations. 
Top-down structure searching is ideal where a low-fidelity model prediction can be validated by a set of DFT capable representative structures.

\section{Background}

\subsection{The fixed-lattice cluster expansion and special quasi-random structures}

Though Ising-like Hamiltonians have been used in physics, chemistry, and materials science for many decades, and the basis used in these Hamiltonians is related to or even used in some atomic structure generation methods.\cite{sanchez_generalized_1984,zunger1990special} The traditional, fixed-lattice cluster expansion was shown to be a spectral method by Juan Sanchez 1984.\cite{sanchez_generalized_1984}. A complete, orthogonal cluster basis may be used to expand functions (e.g. energy) of an alloy on a fixed lattice. 
\begin{equation}
    \begin{split}    
    E_i(\sigma_1, \sigma_2, \cdots, \sigma_{k\in \alpha}) = \sum_\alpha j_\alpha \Phi_\alpha \\
    E_i(\boldsymbol{\sigma}) = \sum_\alpha m_\alpha J_\alpha \bar{\Phi}_\alpha (\boldsymbol{\sigma})  
    \end{split}
    \label{eq:fixed_lattice_ce}
\end{equation}
For example, the energy per atom may be given as a linear expansion in clusters, as in Eq. \eqref{eq:fixed_lattice_ce}.
In fixed-lattice models like this, the only degrees of freedom are the chemical occupations per lattice site.
\begin{equation}
    \sigma_i =
    \begin{cases}
     -m, -m +1,\, \cdots -1, 1, \cdots m+1, m \\
     
     -m, -m +1,\, \cdots -1, 0, 1,\cdots m+1, m \; 
    \end{cases}
    \label{eq:chem_dof}
\end{equation}
The chemical occupations are specified using a chemical variable, $\sigma_i$, for even and odd degrees of freedom per lattice site in Eq. \eqref{eq:chem_dof}.
A complete, orthogonal single site basis is constructed that spans this discrete chemical space per site. 
There is an arbitrary choice for what this site basis may be, but trigonometric bases and discrete Chebychev polynomials are common choices.\cite{sanchez_generalized_1984,aangqvist2019icet,puchala2023casm} A tensor product (cluster) basis is constructed from this site basis,
\begin{equation}
    \Phi_\alpha = \prod_{j \in \alpha} y_m(\sigma_j),
    \label{eq:ce_product_basis}
\end{equation}
where $\alpha$ is the cluster index. In practice, this basis is large. For periodic crystals, the cluster basis is often averaged over the crystal as well as symmetrically equivalent variations of clusters given by space group symmetry operations.
\begin{equation}
    \bar{\Phi}_\alpha(\boldsymbol{\sigma}) = \frac{1}{m_\alpha N} \sum_{\beta = \alpha}\sum_{p}^{N_p} \Phi_\beta(\boldsymbol{\sigma}_\beta(p))
    \label{eq:cluster_corr}
\end{equation}
In Eq. \eqref{eq:cluster_corr}, the cluster functions in Eq. \eqref{eq:ce_product_basis} are summed over points $p$ in the crystal, and the outer sum runs over cluster indices that are symmetrically equivalent to $\alpha$. It is normalized by the symmetry multiplicity of the cluster $m_\alpha$ and the total number of sites $N$. 
The values in Eq. \eqref{eq:cluster_corr} are commonly referred to as the cluster correlations. 
These are evaluated in many cluster expansion codes and used to obtain atomic structures that represent random alloys on fixed lattices.

Special quasi-random structures (SQS) are typically used to represent random alloy phases, a theoretical alloy phase without any chemical correlations, in small atomic structures.\cite{zunger1990special,van2013efficient,aangqvist2019icet} 
How effectively a structure represents a random alloy phase, or any other target system, may be quantified in terms of cluster correlations.
The SQS structures may be defined as those that minimize SQS cost function:
\begin{equation}
    Q_{SQS} = \sum_{\alpha}  |\bar{\Phi}_\alpha(\boldsymbol{\sigma}) - \bar{\Phi}_\alpha^{rand.}| 
    \label{eq:sqs_loss_old}
\end{equation}
where $\bar{\Phi}_\alpha^{rand.}$ are the cluster correlations in a completely random alloy and the sum runs over clusters indexed on $\alpha$ up to some reasonable cutoff. For an equimolar random alloy the cluster correlations are known analytically; for all correlations they are zero, $\bar{\Phi}_\alpha^{rand.}=0$. 
The SQS are generated such that they reproduce the target distribution of cluster functions. A more compact and general notation for the SQS loss function in Eq. \eqref{eq:sqs_loss_old} may be defined as
\begin{equation}
    Q_{SQS} = \Delta \mathcal{M}_1 (d^{s}, d^{t}) .
    \label{eq:sqs_loss}
\end{equation}
By defining the cluster function distribution of the SQS structure $d^s(\{ \Phi_\alpha\})$ and the target distribution as $d^t(\{ \Phi_\alpha\})$, the SQS loss function is just the mean difference between the two distributions. The difference between the first moments, a.k.a. the means $\mathcal{M}_1$, is denoted with $\Delta \mathcal{M}_1$. 
Higher order moments may be added, but for the cases that the target is a completely random alloy, higher order moments such as the variance may be defined analytically in terms of the size of SQS structure used.\cite{zunger1990special} 

Recall that SQS are defined as the structure(s) that minimize the SQS loss function:
\begin{equation}
    \min_{(\boldsymbol{\sigma}_s)} \big[Q_{SQS}\big] .
    \label{eq:sqs_def}
\end{equation}
In Eq. \eqref{eq:sqs_def}, the minimization to find the SQS is performed over the chemical occupations of lattice sites in structure $s$, denoted by $(\boldsymbol{\sigma}_s)$.
In practice, this is done by stochastic sampling algorithms such as simulated annealing (SA).
The initial structures used in SQS are usually a randomly occupied supercells on a parent lattice. Conveniently, the randomly occupied supercells are typically good initial guesses for the SQS solution.

Given the symmetry reduction over space-group operations in Eq. \eqref{eq:cluster_corr}, one may infer that the SQS method becomes increasingly more expensive for low symmetry crystals. This is compounded in materials or alloys with many chemical degrees of freedom per lattice site due to the correspondence between site basis functions, the $y_m$ in Eq. \eqref{eq:ce_product_basis}, and the chemical degrees of freedom in Eq. \eqref{eq:chem_dof}. To generalize SQS-like procedures to arbitrary targets that may be off-lattice or possess large chemical degrees of freedom, it will be important to consider off-lattice generalizations of the fixed-lattice cluster correlations that can handle large chemical degrees of freedom. 


\subsection{The atomic cluster expansion and generalized representative structures}
\subsubsection{ACE}
The ACE method is a natural extension of the traditional fixed-lattice cluster expansion. Though there are many direct parallels, the practically convenient forms are discussed within. Specifically, definitions of the density-projected ACE descriptors will be provided. These descriptors form a complete set of $N$-bond functions of the atomic environment.\cite{drautz_atomic_2019,dusson_atomic_2022,goff2024permutation} 
Given the linear scaling of ACE descriptors with the number of neighbors while still accounting for chemical and spatial degrees of freedom, this complete set of functions to describe atomic environments is a key piece needed to generalize previous structure generation methods such as SQS.\cite{drautz_atomic_2019,lysogorskiy_performant_2021}

The complete set of $N$-bond atomic environment descriptors that, at a high level, is analogous to the fixed-lattice cluster expansion. The construction of ACE descriptors begins with a complete, ideally orthogonal, single bond basis,\cite{drautz_atomic_2019}
\begin{equation}
    \varphi_{i \mu_i \mu_j nlm}(r_{ij},\sigma_{i},\sigma_j) = e_{\mu_i,\mu_j}(\sigma_{i},\sigma_j)R_n(r_{ij}) Y_l^m(\hat{r}_{ij}),
    \label{eq:single_bond}
\end{equation}
where $e_{\mu_i,\mu_j}$ is a chemical basis function that depends on the chemical occupancy of the central atomic site $i$ and the neighbor site $j$, $R_n(r)$ is a radial basis function, and $Y_l^m(\boldsymbol{\hat{r}})$ are the spherical harmonics. Taking the tensor product of these yields the cluster basis,
\begin{equation}
    \{ \Theta_{\mu_i\boldsymbol{{\mu nlm}}}(\boldsymbol{R}) \} = \prod_{\kappa,ij}^\eta  \varphi_{{\mu_i(\mu_j nlm)}_\kappa}(r_{ij}),
    \label{eq:direct_prod}
\end{equation}
where the product is taken over $\eta$ neighbors. 
Though this conceptually may resemble the raw cluster product in Eq. \eqref{eq:ce_product_basis}, it is noted that the computational scaling of this cluster basis goes exponentially with the order of the cluster. 
In practice, It is often more convenient to use the "density trick" to recover linear scaling in descriptor calculations by working with the atomic base:
\begin{equation}
    A_{i \mu_i \mu_j nlm}=\langle \rho | \varphi_{\mu_i \mu_j nlm}(r_{ij}) \rangle .
    \label{eq:projection}
\end{equation}
The atomic density for atom $i$ is projected onto the complete single bond basis \textit{c.f.} Eq. \eqref{eq:single_bond}. The cluster basis in Eq. \eqref{eq:direct_prod} is replaced with a product basis of Eq. \eqref{eq:projection}, yielding:
\begin{equation}
    \boldsymbol{A}_{i \mu_i \boldsymbol{\mu nlm}} = \prod_{\kappa}^N A_{ i\mu_i (\mu_j nlm)_\kappa} .
    \label{eq:atomic_base}
\end{equation}
In Eq. \eqref{eq:atomic_base}, multisets of quantum numbers are used to index the product functions: chemical indices $\boldsymbol{\mu} =(\mu_i,\mu_1,\mu_2,\ldots \mu_N)$, radial indices  $\boldsymbol{n}=(n_1,n_2,\ldots n_N)$, and angular indices $\boldsymbol{l}=(l_1,l_2,\ldots l_N)$ and $\boldsymbol{m}=(m_1,m_2,\ldots m_N)$. Detailed selection rules for these indices and methods to enumerate them may be found elsewhere.\cite{drautz_atomic_2020,dusson_atomic_2022,goff2024permutation}. 
The functions in Eq. \eqref{eq:atomic_base} are invariant with respect to permutations of bonds by construction, but are not invariant with respect to rotations. 

An (over)complete set of rotation and permutation invariant descriptors is obtained by contracting permutation-invariant product functions from Eq. \eqref{eq:atomic_base} with generalized Clebsch-Gordan coefficients.\cite{drautz_atomic_2019}. The generalized Clebsch-Gordan coefficients are commonly used to couple quantum angular momentum states.\cite{yutsis_mathematical_1962} A complete, independent set of ACE descriptors may be obtained from this over-complete set by applying ladder relationships derived for coupling arbitrary quantum angular momentum states when multiset indices in Eq. \eqref{eq:atomic_base} are degenerate,\cite{goff2024permutation,nigam_recursive_2020}
\begin{equation}
 B_{i (\mu_i \boldsymbol{\mu n l L})} = \sum_{\boldsymbol{m}} \mathcal{C}^{\boldsymbol{m}}_{\boldsymbol{l}}(\boldsymbol{L}) \boldsymbol{A}_{i \mu_i \boldsymbol{\mu nlm}} .
 \label{eq:ace_set}
\end{equation}
In Eq. \eqref{eq:ace_set}, the ACE descriptor, in general, gains an additional index multiset, $\boldsymbol{L}$, to explicitly denote which intermediate couplings are used to construct the generalized Clebsch-Gordan coefficients. In Eq. \eqref{eq:ace_set}, the multisets of indices represented with a collected index $\boldsymbol{\nu} = (\mu_i,\mu_1,\mu_2,\ldots \mu_N)(n_1,n_2,\ldots n_N)(l_1,l_2,\ldots l_N)$ $(L_1,L_2,\ldots L_{N-2}, L_R)$. Using this collected index, ACE descriptors will be denoted simply by $B_{i\boldsymbol{\nu}}$. 
For the purposes of structure generation, the final intermediate coupling $L_R$ will be restricted to zero; this corresponds to descriptors that are invariant with respect to rotations. 

The set of ACE descriptors in Eq. \eqref{eq:ace_set} are given for a system with chemical and spatial degrees of freedom, but it may be extended to include other scalar variables, vector-valued variables such as spin and charge transfer, and higher order per-site variables.\cite{drautz_atomic_2020} In this work, we will only consider chemical and spatial degrees of freedom, but the method could be generalized to generate structures with different spin and charge states as well.

\subsubsection{GRS}
The theoretical foundation for the generalized representative structure (GRS) method arises naturally as a generalization of the special quasi-random structure method for fixed lattices. 
The key difference is that GRS relies on with flexible ACE descriptors that are not restricted to fixed lattices. The simplest GRS loss function is defined by replacing the fingerprints in the SQS loss function with ACE descriptors
\begin{equation}
    \begin{split}
    &Q_{GRS} = \alpha_1 \Delta \mathcal{M}_1 [d^s, d^t] + \\
    &\alpha_2 \Delta \mathcal{M}_2 [d^s, d^t] +\cdots ,
    \end{split}
    \label{eq:inter_structure}
\end{equation}
where the generated structure is now characterized in terms of a distribution of continuous, differentiable ACE descriptors $d^s(\{ B_{\boldsymbol{\nu}}\}^s)$ for one structure $s$. 
The target is also characterized with a distribution of ACE descriptors, $d^t(\{ B_{\boldsymbol{\nu}}\}^t)$. This target distribution may correspond to a single atomic structure that need not be on a fixed lattice. 
Higher order contributions, such as the (co)variance are added, denoted with $\mathcal{M}_2$, $\mathcal{M}_3$ for skew, $\mathcal{M}_4$ for kurtosis, and so on.
In practice, the respective contributions to the loss function from the different moments $\mathcal{M}_p$ in Eq. \eqref{eq:inter_structure}, are weighted by a constant $\alpha_p$. 

Sets of GRS may be generated in order to reproduce a target distribution of descriptors. By considering an intra-structure contribution to the GRS loss function, we may ensure that a collection of GRS reproduces moments of some target descriptor distribution,
\begin{equation}
    \begin{split}
    &Q_{GRS}^{intra} = \alpha_1 \Delta \mathcal{M}_1 [d^S, d^t] + \\
    &\alpha_2 \Delta \mathcal{M}_2 [d^S, d^t] +\cdots \, .
    \end{split}
    \label{eq:intra_structure}
\end{equation}
In Eq. \eqref{eq:intra_structure}, the generated descriptor distribution now a joint distribution from $N_s$ structures, $d^S=d^S(\{ B_{\boldsymbol{\nu}}\}^{s_1},\{ B_{\boldsymbol{\nu}}\}^{s_2},\cdots \{ B_{\boldsymbol{\nu}}\}^{s_{N_s}})$. 

For either the inter-structure loss or the intra-structure loss, the target itself may also be comprised of multiple structures to form a joint distribution. A target distribution of $N_t$ target structures will be denoted with, $d^T=d^T(\{ B_{\boldsymbol{\nu}}\}^{t_1},\{ B_{\boldsymbol{\nu}}\}^{t_2},\cdots \{ B_{\boldsymbol{\nu}}\}^{t_{N_t}})$. This is a convenient measure of how close one structure/ensemble of structures is to a second ensemble. 
The utility and accuracy of multi-structure solutions to structure generation loss functions have proven to be useful in other contexts.\cite{karabin2020entropy,sorkin2020generalized} 
Multi-structure solutions can also be obtained with the GRS method to represent complicated distributions that cannot be well represented by a single atomic configuration. This is particularly useful when the target distribution itself is comprised of a set of $N_t$ target configurations with a joint distribution of target fingerprints, $d^T$. 
The methods for generating GRS follow from minimizing the $Q_{GRS}$ loss functions.
\section{Methodology}

\begin{figure}
    \centering
    \includegraphics[width=0.9\linewidth]{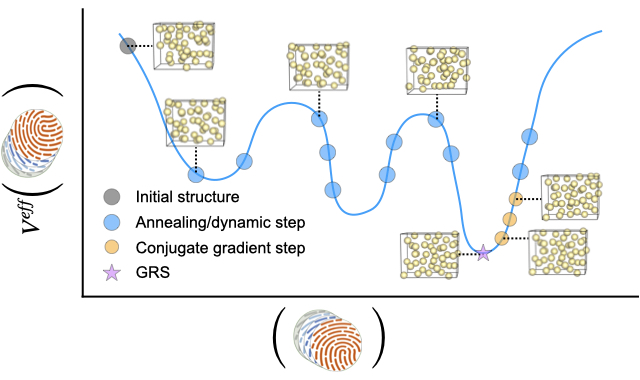}
    \caption{Illustration that highlights the optimization of the GRS loss function for a single atomic structure. An initial structure is annealed on the loss surface are by varying atomic positions and chemical occupations (encoded in the fingerprints on the x-axis). After annealing, a local minimum is found by minimizing the GRS loss function in LAMMPS (the effective potential on the y-axis).}
    \label{fig:min_method}
\end{figure}

A structure represented by atomic descriptors can be compared to any other structure and a quantitative mismatch given in terms of the loss function in Eq. \ref{eq:inter_structure}.
Similarly, a set of structures can be represented with a descriptor distribution, and the mismatch with another descriptor distribution quantified, the loss function in Eq. \eqref{eq:intra_structure}. 
Structures that best reproduce a target distribution are the GRS, and are defined as those that minimize these loss functions. More generally, the summation of Eq. \ref{eq:inter_structure} and Eq. \ref{eq:intra_structure}, with associated weighting terms give the target optimization for GRS, Eq. \ref{eq:grs_set_def}. Generalized representative structures are defined as those obeying
\begin{equation}
    \min_{(R_{s_1}\sigma_{s_1}),(R_{s_2}\sigma_{s_2}), \cdots (R_{s_{N_s}}\sigma_{s_{N_s}})}\big[w_1 Q_{GRS} + w_2 Q_{GRS}^{intra} \big].
    \label{eq:grs_set_def}
\end{equation}
In practice the atomic positions (${R_s}$) and chemical identities (${\sigma_s}$) are not easily intuited, and require optimization until any local or global minima of Eq. \ref{eq:grs_set_def} are to be found. 
The following subsections address each of the key practical concerns that enable GRS generation, followed by practical application of the method.
A pictorial representation of the overall challenge is given in Figure \ref{fig:min_method} where a desired descriptor distribution (purple star) lives at a minima of some effective potential that only depends on the values of the candidate structural and chemical descriptors. 
Proposing an initial candidate structure (grey circle), global optimization steps (blue circles) and local optimization routines (orange circles) that minimize Eq. \ref{eq:grs_set_def} in a computational efficient means are discussed within.

\subsection{Effective potential}
As stated previously, the GRS distinguishes itself from traditional structure optimization methods in that an energy model is not needed \textit{a priori}.
Instead atoms are evolved by utilizing the loss function plus a constraint to avoid close atom contacts,
\begin{equation}
    \begin{split}
    &V^{eff}({\{ \boldsymbol{R}}\}_S,\{\boldsymbol{\sigma}\}_S )=V_{core}({\boldsymbol{R}},\boldsymbol{\sigma})  + w_{1}Q_{GRS}({\boldsymbol{R}_s},\boldsymbol{\sigma}_s ) \\
    & + w_{2}Q_{GRS}^{intra}({\{ \boldsymbol{R}}\}_S,\{\boldsymbol{\sigma}\}_S ).
    \end{split}
    \label{eq:multi_eff}
\end{equation}
Equation \ref{eq:multi_eff} accounts for the interactions among atoms in a single structure, $({\boldsymbol{R}},\boldsymbol{\sigma}$), as well as across multiple structures $\{\boldsymbol{R}\},\{\boldsymbol{\sigma}\}$ that contribute to their respective loss functions $Q_{GRS}$.
The independent variables are used to calculate the ACE descriptors which are smooth and continuous with respect to ${\boldsymbol{R}}$ and $\boldsymbol{\sigma}$, we utilize the efficient LAMMPS implementation of this basis set.\cite{thompson2022lammps,lysogorskiy_performant_2021}

Well-established stochastic methods for sampling chemical degrees of freedom in atomic systems are not efficient for the continuous domain of atomic positions that need to be sampled in GRS methods. 
The key advance enabling optimization of GRS loss functions with dynamic atomic positions is the new ACE capabilities in the machine-learned interatomic potential (ML-IAP) interface in LAMMPS.
This allows a user to define ACE-based interaction models in Python without the need to make compiled language (C++) changes to the LAMMPS code base.
ACE descriptors are passed to the external interaction model, modeled after Eq. \eqref{eq:multi_eff}, where LAMMPS receives per-atom energies and forces needed to perform molecular dynamics, minimization, or any other operations that LAMMPS supports.
Local minima may easily be found on \eqref{eq:multi_eff} by optimizing atomic positions according to conjugate gradient (CG), steepest descent algorithms, and other optimization algorithms used for molecular dynamics.
While a global minimum will help find the best GRS, a key problem will often remains that many target distributions cannot be represented well by a single atomic structure. 
This is especially true if one needs structures near or below 100 atoms for use in electronic structure theory; these may be significantly too small to accurately reproduce complex target distributions of descriptors. 
A set of representative structures may be found by starting with one GRS, adding new candidate structures to the set $S$, and evolving atomic positions such that they minimize the effective potential Eq. \eqref{eq:multi_eff}.

\subsection{Stochastic sampling}
In general, sets of GRS are obtained by combining algorithms that can find global minima with those for local minima.
Depending on the target, favoring either spatial or chemical degrees of freedom or one optimization algorithm over the other may be advantageous. 
For example, targets corresponding atomic structures that lie on (or close to) a fixed parent lattice are sampled with chemical occupation flips or chemical occupation swaps. It is more efficient to find representations of such targets with only SA or genetic algorithms (GA)\cite{forrest1993genetic} rather than including a local optimization method like CG.
Thus, GRS targeting varied local chemistry is performed as it is for fixed-lattices in SQS through chemical occupation flips or swaps.\cite{van2013efficient} 
If atomic positions cannot be fixed based on the target distribution, methods to find local minima are wrapped by the the chemical sampling.
The amount of spatial vs chemical sampling is controlled, in part, through the types of candidate structures proposed, but also through randomly chosen chemical versus spatial sampling. 
The random sampling of chemical versus spatial degrees of freedom is specified through: (1) the probability of moving and/or optimizing atomic positions through iterative optimizers and (2) the probability for sampling new chemical occupations through chemical occupation flips, swaps, or a different candidate structure. 
Where the parent lattice is known, or a single chemistry is desired, one can nullify one of these probabilities.

For the GA, two different crossover functions may be used. One crossover function combines a fraction of one structures atomic coordinates and chemical occupations with the coordinates and chemical occupations of a second structure. 
This is accomplished by collecting a random slice of atoms from two parent structures and combining them in one of the parent cells. If this results in atomic overlap, one of the overlapping atoms is removed. The second crossover function used takes the atomic coordinates from only one parent, but allows for mixing of the chemical occupations.
From a given generation of atomic structures, candidates were selected using a tournament selection method. The GRS loss function is used directly in the tournament selection method to evaluate the fitness of the candidates.

For the SA optimizer, the allowed steps are specified based on the target distribution (e.g. chemical vs spatial). Beyond these steps that were defined earlier along with the GA "mutations", the simulated annealing method is characterized by the acceptance criteria. In this case, the acceptance criteria is defined in terms of differences in the GRS loss function from one Monte Carlo step and the previous step as
\begin{equation}
    \mathrm{argmin} \bigg(1,  \;\; exp\big(-\beta [Q_{GRS}^1 - Q_{GRS}^0 ]\big) \bigg).
    \label{eq:sa_criteria}
\end{equation}
In Eq. \eqref{eq:sa_criteria}, the acceptance criteria for SA is given in terms of the GRS loss function at different Monte Carlo steps and a thermodynamic beta, $\beta=1/k_b T$ that artificially introduces temperature. 
The beta is initialized such that it corresponds to an large initial temperature. After a sufficient number of steps, it is decreased.

Either the SA or the GA is used within for global optimization when necessary. 
For the majority of the applications the GA is applied at least a single generation of atomic structures from the GA is used for parallel sampling.
In some cases, accurate representative structures may still be obtained using only the CG method in LAMMPS, provided that appropriate candidate structures are proposed.

\subsection{Initial conditions}
Given the broad range of target distributions accessible with GRS, it is important to have good initial guesses for structures. If not, reasonable minima may not be found. 
For comparison, the residual of the SQS loss function (the minimal value of the SQS loss function) is related to cell size and the target distribution and initial candidate structures are chosen with some consideration of the target .\cite{van2009multicomponent,van2013efficient,aangqvist2019icet,zunger1990special} 
When the target in SQS is a random alloy, a randomly occupied supercell is often used as an initial candidate structure. More accurate representations of the random alloy target may be obtained with larger supercells.

Currently, candidate structures in GRS are also proposed according to the target distribution of ACE descriptors. 
Target distributions of ACE descriptors corresponding to fixed-lattice systems are assigned with random atomic positions close to those from the parent lattice. 
The parent lattice may be identified through an appropriate ACE descriptor set or, if a single atomic structure is set as the target, by atomic structure symmetry libraries.\cite{togo2024spglib}
Liquid and amorphous systems are assigned randomly occupied supercells with variable density close to those sampled by the target.
Chemical occupations are assigned randomly and according to restrictions based on global composition.
If more flexibility in spatial degrees of freedom is required, supercells of arbitrary parent lattices may be generated using Hermite normal form transformations, similar to the first step in Ref.~[\citenum{morgan2017generating}].
More thorough spanning of chemical occupations may be accessed by using fixed-lattice candidates generated with existing algorithms, or similar methods to span unique chemical labelings of supercells.\cite{morgan2017generating,aangqvist2019icet}  
The classes of initial candidate structures used presently for GRS are obviously not exhaustive of all possible atomic configurations. However, as is demonstrated in our results, the current variability in candidate structures is sufficient for starting the GRS optimizations with diverse solid-state targets.
Candidates relevant for molecular and organic systems could easily be introduced using analogous arguments for domain expertise and symmetry. However, the initial candidate structure types already presented may be useful for some coarse-grained molecular systems.

Lastly, the initial conditions of the ACE basis set that underpin all of these sampling methods needs to be defined. 
The ACE descriptor sets in theory are complete, but in practice the body order needs to be truncated much like the fixed-lattice cluster functions. 
Unlike the fixed-lattice cluster functions, the radial and angular basis functions used to construct ACE descriptors need to be truncated per body order. 
Additionally, hyper-parameters for the radial basis need to be specified; by default with the LAMMPS interface, this requires specifying hyper-parameters for the radial basis used in Ref.~\citenum{drautz_atomic_2019}. 
To truncate the body order of the ACE descriptor sets, complete sets of ACE descriptors up to 5-body (rank 4) are enumerated.\cite{goff2024permutation} 
The maximum radial function indices per descriptor rank are: $n_{max}=8,2,2,1$ and angular $l_{max}=0,2,2,1$.
For the 4-element Fe-Cr-Si-V system, the same set is used, but it is truncated at 4-body.
The radial function cutoffs are specified based on tabulated van der Waals radii.\cite{larsen2017atomic} 
For most relevant condensed matter systems considered within, these cutoffs correspond to 4-7 neighbor shells.
The scaled distance parameters, $\lambda$ in Ref.~\citenum{drautz_atomic_2019}, were specified to be 5\% of the radial cutoff. 
The efficacy of many other values of radial cutoffs and scaled distance parameters were tested for single and multi-element target systems. 
The 5 \% fraction was chosen because, regardless of radial cutoff, it showed improved information content compared to other hyper-parameter choices with lower global feature reconstruction errors.\cite{goscinski2023scikit} 
Similar tests for radial cutoffs and higher body order ACE descriptors show that information content is already saturated with the values used for the example systems tested. This helps show that the descriptor sets used are sufficient for these examples. It is important to note that the saturation of information content is dependent on the target systems. 
It has been rigorously shown that some motifs and structures cannot be represented with low body-order functions and short cutoffs, so ACE bases should be truncated carefully.\cite{pozdnyakov2020incompleteness,nigam2024completeness}
It is noted that the parameters corresponding to an optimized ACE interatomic potential may differ.\cite{bochkarev_efficient_2022}

\section{Results}

\subsection{Representing chemical ordering}
The first result presented for GRS leverages the relation between the ACE descriptors and fixed-lattice cluster correlations to produce fixed-lattice structures.
Recent SQS implementations allow one to specify the targeted cluster correlations; these correlations may correspond to an ordered alloy.\cite{van2013efficient,aangqvist2019icet} 
Where the solution is known exactly and can be produced with previous fixed-lattice methods is an ideal test case for GRS.
For this demonstration, the ACE descriptors (up to 5-body) for the $L_{1_2}$ intermetallic of Au-Cu were set as a target distribution, a single GRS is sought to match this. 
The different stochastic optimization algorithms, the GA and SA, were assessed while reproducing this ordered intermetallic with respect to random walks for comparison. 
Simulated annealing was performed with two fictitious temperatures $\beta = \{0.01,2\}$.
In this example, steps were restricted to be chemical occupation flips on a fixed lattice. 
For the GA, a maximum of 80 generations were allowed with 50 structures per generation. 
The mutation and crossover probabilities for the GA were 0.6 and 0.4, respectively.\cite{forrest1993genetic} 
Initial structures for all three sampling methods were $(2\times 2 \times2)$ cubic FCC supercells with randomized Au and Cu occupations. 
With simulated annealing and the random walks, a maximum of 5000 steps was allowed per temperature. 
This is a reasonable starting point to sample the chemical occupations of the 32 atomic sites in the $(2\times 2 \times2)$ supercell.

\begin{figure}
    \centering
    \includegraphics[width=0.8\linewidth]{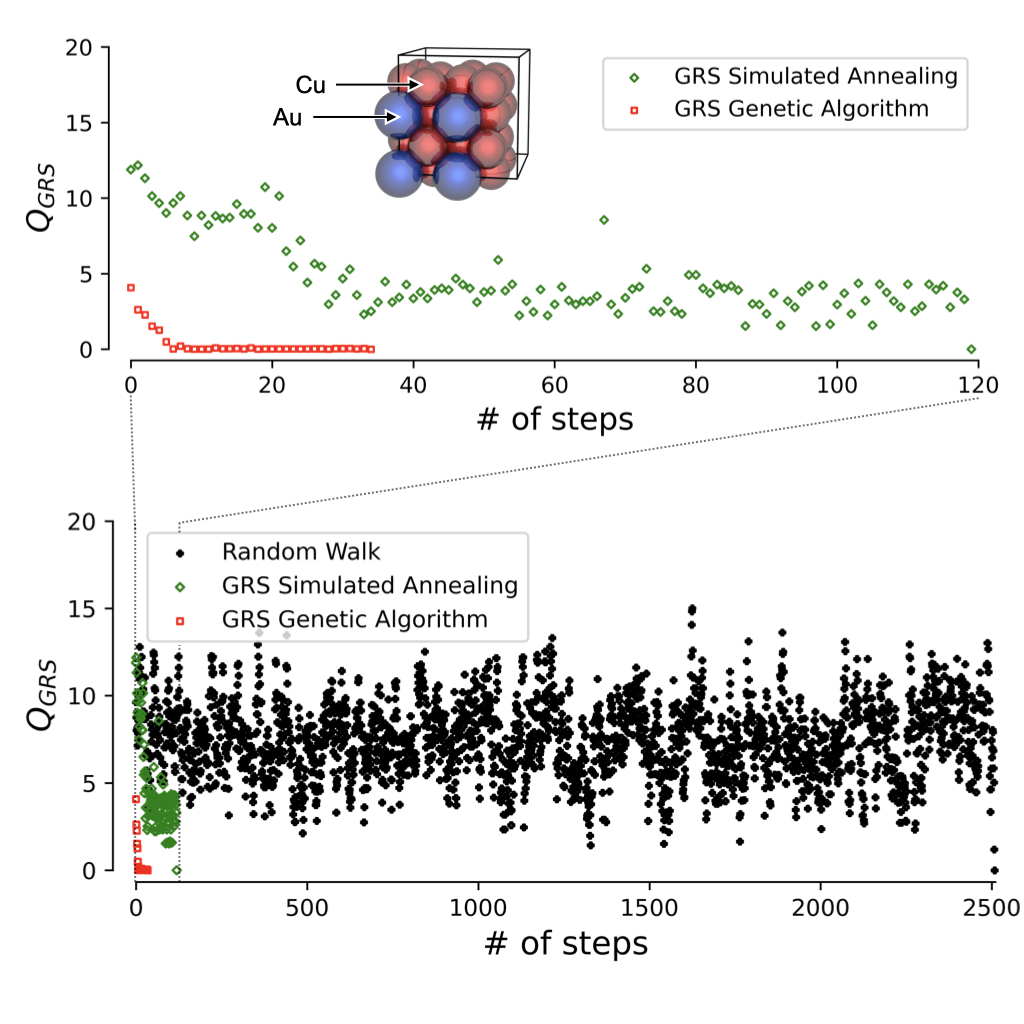}
    \caption{The number of iterations in needed by GRS simulated annealing (green plus symbols), GRS genetic algorithm, and random walks (black diamonds) to reproduce the simple $L_{1_2}$ intermetallic AuCu$_3$ exactly in a $2\times2\times2$ supercell (pictured).}
    \label{fig:grs_vs_rndwlk}
\end{figure}

In Fig. \ref{fig:grs_vs_rndwlk}, it can be seen that the GRS method exactly reproduces the intermetallic with $Q_{GRS}=0$, significantly more rapidly than random walks. Random walks reached an exact representation of the intermetallic after 2518 steps, but this is for a relatively simple crystal structure to reproduce. 
For comparison, the GRS method using simulated annealing to optimize the loss function reproduced the structure in 120 steps. The GRS method using the genetic algorithm to optimize the loss function reproduces the structure exactly in 35 generations. 
From this vastly improved convergence to a solution for this simple system, it is evident that GRS method offers another approach for generating crystal structures with specified chemical order (or lack thereof). 
This is the same result obtained with SQS methods that allow one to specify the desired chemical order on a parent lattice.\cite{aangqvist2019icet}
However, since this is done through the ACE descriptors in GRS, this procedure is not limited to fixed parent lattices. Through GRS, specified chemical ordering or disordering may be targeted on more exotic structures including polymer chains, molecules, and amorphous materials.

\subsection{Non-crystalline structures}

\begin{figure}
    \centering
    \includegraphics[width=0.9\linewidth]{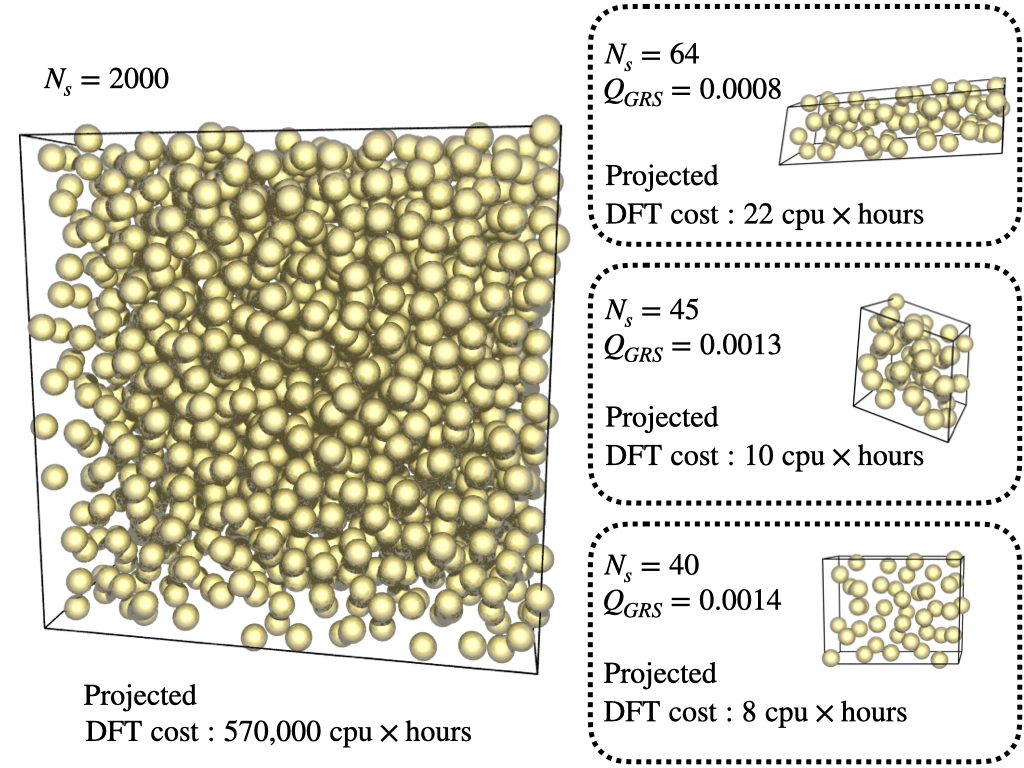}
    \caption{The GRS generated with different numbers of atoms, $N$. These reproduce the ACE descriptor distributions of a liquid Ta target system with increasing accuracy, lower minimum value of $Q_{GRS}$. For each GRS, the projected DFT cost is provided according to an $N^3$ scaling rule derived from Quantum Espresso self-consistent field calculations.\cite{giannozzi2009quantum} }
    \label{fig:ta_liq}
\end{figure}

Moving to a demonstration of arbitrary structural complexity, the capability to represent amorphous and liquid structures with GRS was also assessed. 
Utilizing a pre-existing interatomic potential from Ref. \citenum{thompson_spectral_2015} was used in MD simulations of liquid Ta at 4000 K. 
A snapshot of a 2000-atom system after it was equilibrated was used as a target. 
With the goal of representing the melt in a single DFT-sized structure, single GRS structures were generated with different sizes and the accuracy with which the GRS structure reproduces the liquid Ta system reported, presented in Fig. \ref{fig:ta_liq}. 
The trend continues that larger GRS structures more accurately represent the mean descriptor values of the targets as well as the variance in the descriptors of the target. 
This demonstrates that GRS method can be used to make representative liquid an amorphous states with systematically improvable accuracy. 
Extensions of the displayed methods are the ability to validate MD predictions (up to the representation accuracy) using higher fidelity methods.
This top-down multiscale modeling capability has the ability to address problems where computational cost of a purely DFT prediction is prohibitive.

Reported in Fig. \ref{fig:ta_liq} is the reduction in DFT time by constructing a minimal size GRS. 
Similar to how SQS makes DFT calculations for complex alloys more computationally tractable, the GRS method makes DFT calculations more tractable for liquids and other systems as well. Structures that more or less accurately reproduce a target system / target fingerprint distribution may be generated such that subsequent DFT calculations fit a computational budget. 
The computational cost for single SCF calculations in Quantum Espresso are used to predict the cost of the generated GRS structures of liquid Ta as well as the Ta melt system itself (assuming an $N^3$ scaling).\cite{giannozzi2009quantum} 
As expected the computational cost is significantly lower than DFT for the larger melt, however they retain much of the same information, as reported by the $Q_{GRS}$ residuals for these structures.
$Q_{GRS}$ residual is the value of the loss function remaining after minimization.
This quantifies (1) the mismatch between the average descriptor values of the target system and the GRS and (2) the mismatch between the variance in the descriptor values between the target system and the GRS according to Eq. \eqref{eq:inter_structure}. 
For the values reported in Fig. \ref{fig:ta_liq}, the weight for each term in Eq. \eqref{eq:inter_structure} is one. 
These results indicate that not only does the GRS method produce representations of larger structures that reproduce the same atomic environments, but also how varied the atomic environments are. 


\subsection{Reducing structure ensembles}
To further exercise the GRS method, it is useful to demonstrate the capability to generate sets of GRS that retain the information content from a larger ensembles of structures.
Here, the GRS method is used to reproduce a training dataset used for an MLIP, the same that was used to generate the liquid structure\cite{thompson_spectral_2015}. 
Due to the domain expertise of the MLIP developer, these datasets do tend to be centered on the ground-state or low-energy structures. 
Recent works have demonstrated key importance of including some high-energy structures produced from ab-initio MD and other methods, but proportionally there still tend to be more low energy structures.\cite{sikorski2023machine} 
As a result, and as observed in most training workflows, much of the atomic environment information is redundant.\cite{rohskopf2023fitsnap} 
In other words, a lot of low-energy structures have similar descriptors. 
To simplify MLIP training, and reduce the computational cost of DFT used in these workflows, it may be desirable to reduce the number of low-energy structures. 
For these reasons, GRS sets were developed to reproduce the descriptor distribution of a known dataset. 
This was done by finding sets of structures that satisfy the GRS criteria, for a target distribution of ACE descriptors generated by the 373 structures the MLIP training dataset, $d^T(\{ B_{\boldsymbol{\nu}}\}^{t_1},\{ B_{\boldsymbol{\nu}}\}^{t_2},\cdots \{ B_{\boldsymbol{\nu}}\}^{t_{373}})$. 

\begin{figure}
    \centering
    \includegraphics[width=0.8\linewidth]{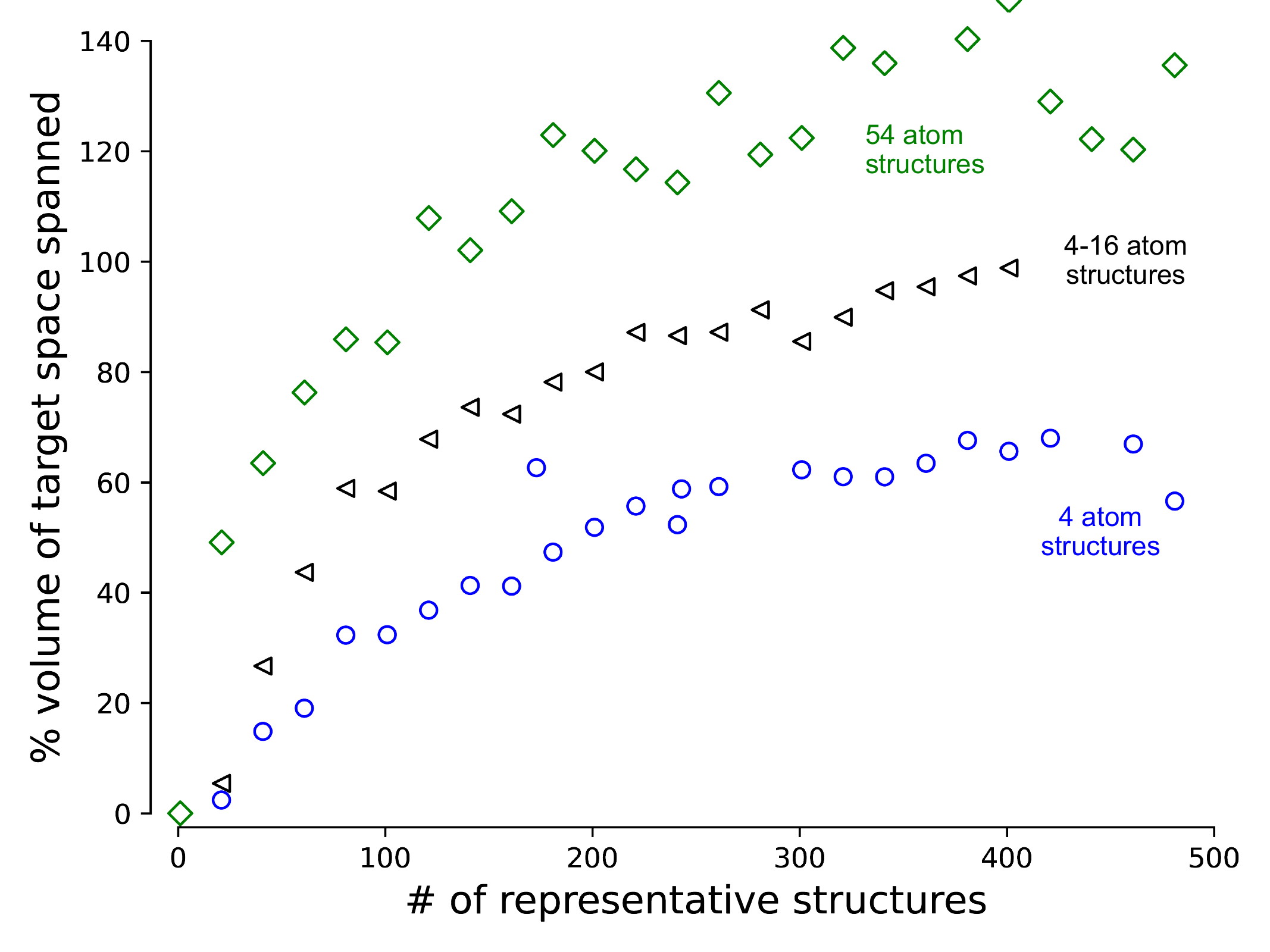}
    \caption{The number of GRS configurations needed to span the same ACE descriptor space volume as the Ta dataset in Ref. \citenum{thompson_spectral_2015} for different representative structure sizes.}
    \label{fig:convergence}
\end{figure}

The reported Ta training dataset has 373 configurations in it, with a total of 4,224 atom environments\cite{thompson_spectral_2015}. 
Here we aim to demonstrate the varying degrees of accuracy with which a known dataset may be reproduced with a set of GRS, and if this may be used to reduce the total number of structures or atomic environments needed to retain the same information.
In Fig. \ref{fig:convergence}, sets of GRS with different numbers of structures and/or different numbers of atoms per structure are used to reproduce the Ta dataset. 
The maximum number of atoms per structure generate different data series. 
The number of structures in the GRS set are given as the independent variable, while the dependent variable is a measure of the accuracy with which the target is reproduced. 
Principal Component Analysis (PCA) of descriptors (in dimensions that reaches 99\% explained variance) conveniently compares the descriptor distributions of the GRS sets and the target as it directly captures the desired variance of the dataset.
\begin{equation}
    V^{Rep} = V_{GRS}^{hull} / V_{target}^{hull}
    \label{eq:vol}
\end{equation}
Convex hulls are generated in the PCA space of the target distribution for both the GRS and the target and the volumes of the convex hulls are compared.
The metric in Eq. \eqref{eq:vol}, the relative volume of the GRS sets descriptor distribution is compared to the descriptor distribution of the target. 

As was also seen in Fig. \ref{fig:ta_liq}, larger structures more accurately represent the complicated target distribution of the MLIP training dataset.
Additionally, allowing for more structures produces more accurate representations of the target distribution.
The GRS set comprised of larger 54-atom structures more rapidly converges to 100 \% of the volume of the descriptor space. 
This result follows intuitively from what was first reported for traditional SQS methods, where the completely random alloys were best represented by larger SQS, and smaller SQS structures had larger residuals.\cite{zunger1990special} 
The GRS set comprised of smaller (e.g. 4-atom) structures does not entirely reproduce the target distribution of the MLIP dataset with 500 structures. 
However, it should be noted that at maximum this is comprised of 2000 atomic environments reaching upwards of 60 \% of the descriptor space volume with less than half of the total atomic environments of the target.
It may be possible to reach 100 \% of the targets descriptor space volume with only 4-atom structures, however the inter-structure residuals,  $\Delta \mathcal{M}_1(\boldsymbol{R}_s), \Delta \mathcal{M}_2(\boldsymbol{R}_s)$, within any one 4-atom representative structure may be high. 
If inter-structure residuals are too high, it may be difficult to represent the target accurately. 
To show that the $V^{Rep}$ is not yielding a false success due to sampling a large volume of descriptor space completely unlike the target, histograms of the descriptors are also reported for both the target distribution and the generated GRS sets in the supporting information SI Fig. 1.

\subsection{Unsupervised structure generation}

Brute-force exhaustive enumeration of crystal structures with large chemical degrees of freedom can be expensive. 
Instead a variation of the GRS method can be used to enumerate atomic structures that are \textit{dissimilar} to a target system. 
This is particularly useful for sampling chemically complex materials. 
If a random alloy is set as a target, one can systematically enumerate structures that are further from the random alloy both chemically and spatially with a tuneable strength for each chemical and spatial degrees of freedom, see Eq. \ref{eq:grs_set_def}.
This may be done simply by flipping the sign of $\mathcal{M}_1$ in Eq. \eqref{eq:inter_structure} and adjusting probabilities for chemical versus spatial sampling. 
This method for reversing the sign of the $\mathcal{M}_1$ term in the GRS loss function may be referred to as the reverse GRS method.

\begin{figure*}
    \centering
    \includegraphics[width=0.7\linewidth]{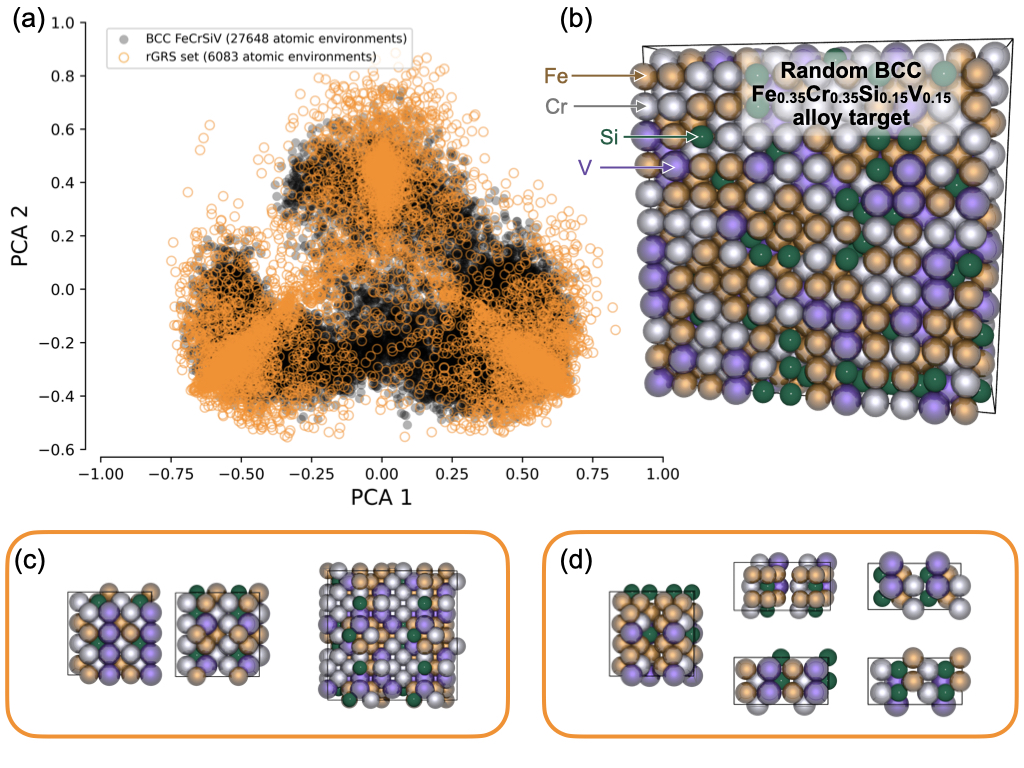}
    \caption{(a) The combined PCA analysis comparing a set of GRS generated to find structures \textit{dissimilar} to a random alloy target, Fe$_{0.35}$Cr$_{0.35}$Si$_{0.15}$V$_{0.15}$, system (b). This generates crystal structures with larger chemical correlations than the target (c) and larger structural distortions than the target (d). For (c) and (d), the some GRS are visualized as $2\times 2\times2$ supercells to show periodicity. }
    \label{fig:fecrsiv}
\end{figure*}

In Fig. \ref{fig:fecrsiv} some results for enumerating new atomic structures structures that may be relevant for modelling BCC to FCC phase transitions in a Fe$_{0.35}$Cr$_{0.35}$Si$_{0.15}$V$_{0.15}$ alloy. 
Here $w_1$ and $w_2$ were set to -0.01 in Eq. \eqref{eq:multi_eff}. 
This weakly pushes the representative structures away, as indicated with the PCA analysis in Fig. \ref{fig:fecrsiv}(a), from the random alloy target system visualized in Fig. \ref{fig:fecrsiv}(b). 
This demonstrates the ability to enumerate atomic structures that have systematically decreasing similarity with a target. 
Keeping the dissimilarity weak allows one to generate crystal structures (e.g. BCC, FCC) with more chemical correlations / more chemical ordering than the randomly ordered target, Fig. \ref{fig:fecrsiv}(c).
Along with producing structures with more chemical ordering than the target, the reverse GRS method produces structures with atomic positions dissimilar to the perfect BCC parent lattice of the target, Fig. \ref{fig:fecrsiv}(d).

This result demonstrates key utility in unsupervised structure generation for atomic systems with chemical and spatial degrees of freedom. 
Rather than generating atomic structures based on thermodynamic stability, which may rely on imperfect energy models, structures are generated based on change in a complete set of atomic environment descriptors. 
If a target is set that is expected to be favorable based on chemical and physical intuition, reverse GRS will systematically step outward from this structure in the space of chemical occupations and positions of atomic sites. 
As new structures are generated for a given $w_1$ and $w_2$ pair, reverse GRS can then bootstrap to unique structures by redefining the target. 
This is a more informed search than brute-force structure enumeration based on symmetry or information content, and does not require DFT calculations for each proposed structure to be evaluated.  
Additionally, the amount of new information or new atomic environments is an adjustable parameter, which is a unique tool for users that are also developing MLIP.
Further exploration of the unsupervised structure searching in the chemical occupations or atomic positions relative to some target system is given in the supporting information. 
There it is shown how the GRS method can, with an increased amount of chemical sampling, span chemical environments generated by exhaustive atomic structure enumeration methods for fixed lattices.\cite{morgan2017generating} 
Depending on the number of structures allowed in the representative set, this may come at a cost of decreased sampling of spatial degrees of freedom. 
Increased chemical sampling in an set may result in poorly capturing thermal variations in atomic positions in ab-initio molecular dynamics (AIMD).
This may be remedied by increasing spatial sampling where more of the atomic environments spanned by AIMD may be accessed using GRS in this unsupervised approach. 
These results suggest that the GRS method may be used to sample both symmetrically distinct atomic structures and spatially varied atomic structures simultaneously with a degree of tuneability that was not possible with previous methods.

\subsection{Reducing inhomogeneous structures}

The GRS method was also applied to represent large atomic structures with complex and varied atomic environments. 
An example is made in the representation of atomic structures corresponding to rare, collective events in molecular dynamics simulations.
While snapshots of large MD simulations at rare events and other large complex atomic systems are obviously not well-represented in a single <100 atom structure, a set of atomic structures may collectively represent large complex atomic systems more effectively. 
As with other applications of GRS, this may be done with systematically improvable accuracy and resolution.
\begin{figure}
    \centering
    \includegraphics[width=0.8\linewidth]{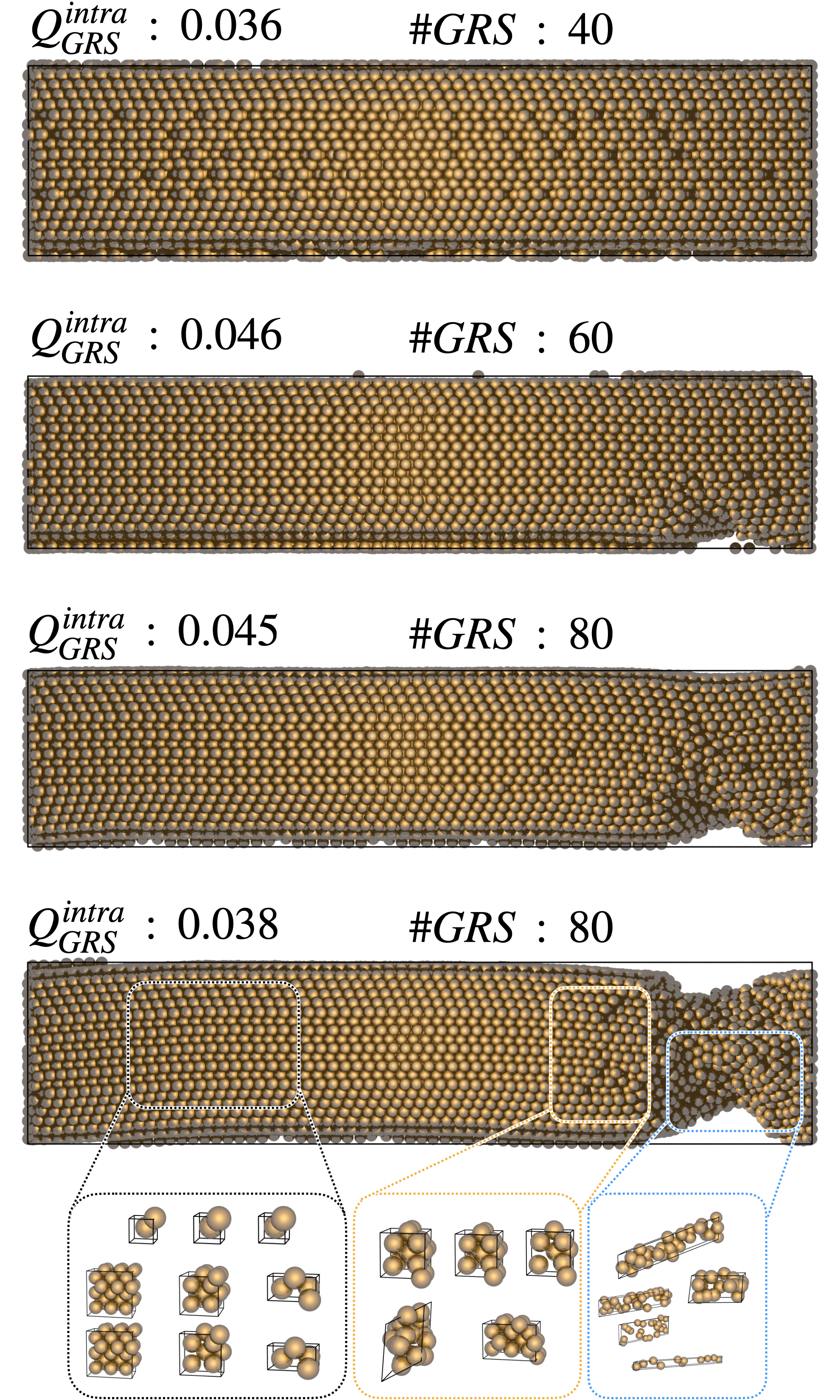}
    \caption{BCC Iron tensile test using an EAM interatomic potential, the loading direction is along [111] in a cylindrical wire geometry. Snapshot is post mechanical yielding showing significant necking (diameter reduction) and corresponding non-crystalline regions where plastic deformation has occurred. }
    \label{fig:repro}
\end{figure}

For the sake of argument, one could generate a complex result from a MD simulation from a low-fidelity interatomic potential with the aim to produce a quantum accurate MLIP to improve these predictions. 
This complex structure, and its corresponding descriptor distribution, may not be identical to the prediction generated from a higher fidelity model, but will occupy a similar region of descriptor space.
If in fact this low-fidelity descriptor distribution does not occupy the desired high-fidelity descriptor space, one could employ bootstrapping or active-learning methods until the desired accuracy is achieved.
The accuracy of any given interatomic potential is not the focus of this work, but consider the computationally efficient Embedded Atom Method \cite{daw1993embedded} as the low fidelity model form of a Fe-Cr alloy\cite{eich2015embedded}.
Figure \ref{fig:repro} captures a tensile test of an initially defect-free BCC Fe wire with a 3.6 nm cross-section, strain direction aligned to the [111] crystal axis.
At the onset of plasticity new atomic environments are generated around slip planes, and in large number where necking (diameter reduction) is present. 
For snapshots along the trajectory of the tensile test, the GRS method was employed to reproduce the information contained in the atomic structure. In Fig. \ref{fig:repro}, the number of structures needed to span the descriptor space volume is reported along with the MAE in reproducing the target distribution of descriptors is reported. 
Before necking, fewer GRS ($\simeq$40) are needed to span the same volume in descriptor space. The final state in the simulation requires more GRS instances. For this final state, some of the GRS are also pictured to show which areas of the simulation the representative structures bear similarity with. 
There are crystalline GRS that correspond to regions further from the necking, distorted crystalline structures closer to the necking, and surface-like and amorphous structures that bear the most resemblance to the region with necking. These generated structures are sizes actually accessible with DFT, unlike the target itself. 

Apart from the qualitative similarity of some structures with different regions in the MD snapshots for Fig. \ref{fig:repro}, quantitative similarity may also be assessed.
Quantitatively, the quality of the set of GRS structures may be assessed using straightforward metrics such as the mean average difference between the first and second moments of the  distribution of ACE descriptors for the MD snapshot (the target) and the set of GRS. A value closer to 0 indicates a closer match to the target distribution. It is important to note that these metrics are not normalized in the same way that residuals are for the SQS methods. This same metric measured for well-converged SQS structures (structures that closely match the targeted random alloy distribution of ACE fingerprints) have scores in the range 0.1-0.2. 


The ability to distill the atomic environment information in complex MD systems and large-scale simulations offers a valuable tool for active learning and interatomic potential development. It can help more generally in multi-scale studies of materials; large-scale systems may be downsized for direct calculations in DFT. This is limited by the number of structures needed to reproduce the large systems. In the case above, 80 GRS structures were used to reach a reasonable reproduction of the MD snapshot.



\section{\label{conclusions}Conclusions}
A true generalization of structure generation methods is achieved through the use of atomic descriptors that encode all necessary degrees of freedom. 
Since ACE descriptors are continuous, differentiable, and span both chemical and spatial degrees of freedom for atomic environments, the GRS method is not restricted to fixed lattices or high-symmetry crystals. 
Furthermore, the extension of the GRS method to be utilized as an effective interaction potential between atoms allows for a wide array of applications for dimensional reduction of large data sets, solution of MD to DFT validation of complex phenomena, and unsupervised generation of dissimilar structures. 
While the effective potential constructed between candidate and target ACE descriptors is not physically intuitive, we have shown that it can achieve dramatic reduction of computational cost of purely DFT structure search methods. 
Moreover, as employed the GRS effective potentials utilize efficient ACE descriptors calculated in LAMMPS, which also enables users to capitalize off the diverse functionality of this code base. 
The only input required for the GRS method is a distribution of descriptors as the target, which can be initialized from known atomic structures. 
In addition to high-symmetry crystals, the input atomic structures may be low-symmetry systems.
The GRS method is capable of accurately representing low-symmetry systems as well as high-symmetry crystals with systematically improvable accuracy. It was shown for an archetypal example that GRS can reproduce some atomic structures exactly.
One compelling result presented is that the GRS method is able to produce a super-set of structures one would obtain from fixed-lattice structure generation methods like SQS. 

More exact connections can be drawn between SQS and GRS if the density trick is not used in the ACE descriptors. Doing so would turn the $\mathcal{O} (N) $ scaling into $\mathcal{O} (N^x) $ where $x$ is the body order of the ACE clusters. This resulting lack of performance could reduce the applicability of this method in large-scale systems, including the MD snapshots demonstrated.
The ability to represent large MD snapshots with multiple small DFT-compatible cells is useful for validating MD results and for active learning workflows.
Both the training dataset reduction and reproducing large MD snapshots of rare events demonstrate the utility of the GRS method for data reduction. It is also demonstrated that GRS may be used to sample new atomic configurations with specific structural and chemical motifs. This capability of GRS to represent targeted structural and chemical motifs could be used to drive simulations towards rare or important events such as reactions and transition states. 
Targeting transition states and other rare events in large-scale atomic systems with the GRS method should be a key focus in future studies.

\begin{acknowledgments}

Goff, Mullen, Starovoytov, and Yang gratefully acknowledge funding support from U.S. Department of Energy, National Nuclear Security Administration (NNSA) under MSIPP NSAM-ML Consortium under contract DE-NA0004112.
Goff, Mullen, and Wood gratefully acknowledge funding support from the U.S. Department of Energy, Office of Fusion Energy Sciences (OFES) under Field Work Proposal Number 20-023149.
Starovoytov, and Yang gratefully acknowledge funding support from DOE/NNSA award No. DE-NA0004144.
Starovoytov, and Yang gratefully acknowledge funding support from NSF award No. 2216805, 2019561, 1915520, 2154344, OAI-2019511, OAI-1946231, NSF(2024)-SURE-302, USDA award No. 2024-38427-43463.
We acknowledge support from LONI providing computational resources, loni\_mat\_bio20 allocation, with technical support from Phillip Marr.
Sandia National Laboratories is a multi-mission laboratory managed and operated by National Technology and Engineering Solutions of Sandia, LLC, a wholly owned subsidiary of Honeywell International, Inc., for the U.S. Department of Energy's National Nuclear Security Administration under contract DE-NA0003525. 
This paper describes objective technical results and analysis. Any subjective views or opinions that might be expressed in the paper do not necessarily represent the views of the U.S. Department of Energy or the United States Government.
\end{acknowledgments}

\bibliography{main.bib}

\clearpage
\newpage
\onecolumngrid

\end{document}


\title{Supporting Information: Generalized representative structures for atomistic systems}

\author{James M. Goff}
\affiliation{Center for Computing Research, Sandia National Laboratories, Albuquerque, New Mexico 87185, USA}
\author{Coreen Mullen}
\affiliation{Center for Computing Research, Sandia National Laboratories, Albuquerque, New Mexico 87185, USA}
\affiliation{Department of Mathematics, Computer Science, and Engineering Technology, Elizabeth City State University, Elizabeth City, North Carolina 27909, USA}
\author{Shizhong Yang}
\affiliation{Computer Science Department, Southern University, Baton Rouge, Louisiana 70807, USA}
\author{Oleg N. Starovoytov}
\affiliation{Computer Science Department, Southern University, Baton Rouge, Louisiana 70807, USA}
\affiliation{Center for Computation and Technology, Louisiana State University, Baton Rouge, Louisiana 70803, USA}

\author{M. A. Wood}
\affiliation{Center for Computing Research, Sandia National Laboratories, Albuquerque, New Mexico 87185, USA}

\maketitle

\clearpage
\newpage
\onecolumngrid
\section{Validating descriptor distribution for GRS sets}
For Fig. 5 in the main text, a metric was used to compare the volume of the ACE descriptor space spanned by the GRS sets compared to the volume of descriptor space spanned by the MLIP training dataset used as a target. 
This metric did not give information how well the values of descriptors from the GRS sets compared to the target. 
Here we confirm that they are similar with some more detailed analysis of the descriptor distributions of the GRS sets generated to represent the Ta MLIP training dataset.\cite{thompson_spectral_2015}

\begin{figure}[H]
    \centering
    \includegraphics[width=0.5\linewidth]{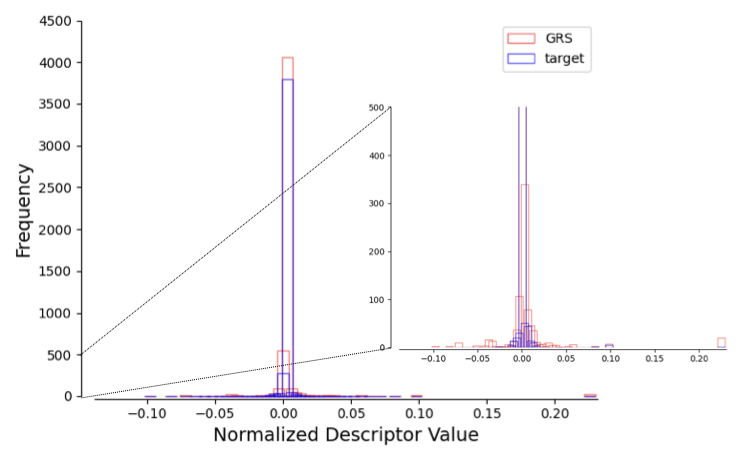}
    \caption{The descriptors sampled by the GRS set (red) compared to the target distributions (blue). These normalized histograms demonstrate that the GRS set samples many of the same descriptor values as the targeted dataset.}
    \label{fig:main_histogram}
\end{figure}

In Fig. \ref{fig:main_histogram}, a histogram of descriptor values in the GRS set (red) is compared to the descriptor values sampled by the original Ta training dataset Ref. ~\citenum{thompson_spectral_2015}. The comparison indicates that the descriptor values sampled by the generated GRS structures closely match those in the Ta dataset. While this is only one descriptor of 15 used to generate the GRS set for the Ta dataset, the others follow similar trends in how closely the GRS set matches the Ta dataset. 
These histograms comparing the descriptor values sampled help confirm that not only are the GRS sets representing a similar volume of descriptor space as the target, but the values of descriptors are similar. 

\section{Comparing novel structure generation with GRS to other methods }

The sets of representative structures methods may also span compositions in alloys and multi-component systems like fixed-lattice structure generation methods do. However, an advantage of the generalized representative structure methods is that sets of structures may be generated that span compositions, but also varied spatial distributions of atoms. 
To demonstrate this flexibility, GRS sets are compared to other state-of-the-art structure generation methods, particularly those used in alloy models and MLIP training.\cite{aangqvist2019icet,sikorski2023machine} For comparison with alloy model datasets, all symmetrically distinct alloy structures for Fe-Cr were generated for FCC and BCC parent lattices up to a cell size of 6.\cite{morgan2017generating,aangqvist2019icet}
This resulted in 274 symmetrically distinct structures. Sometimes these sets of structures are referred to as symmetrically inequivalent supercells (SIS), and these exhaustive generate all distinct chemical occupations on a fixed lattice (with some degree of coarseness based on the maximum supercell size). Performing Ab-initio molecular dynamics (AIMD) on SIS (or the 1-element equivalent) or other supercells are key components of training datasets for machine-learned interatomic potentials.\cite{meziere2023accelerating}
As SIS sets and AIMD are the current state-of-the-art enumeration methods used for alloy sampling and MLIP training in the computational materials community, both serve as excellent references for comparison for the newly developed GRS method. 
This comparison was made for an Fe-Cr binary alloy.
\begin{figure}
    \centering
    \includegraphics[width=0.65\linewidth]{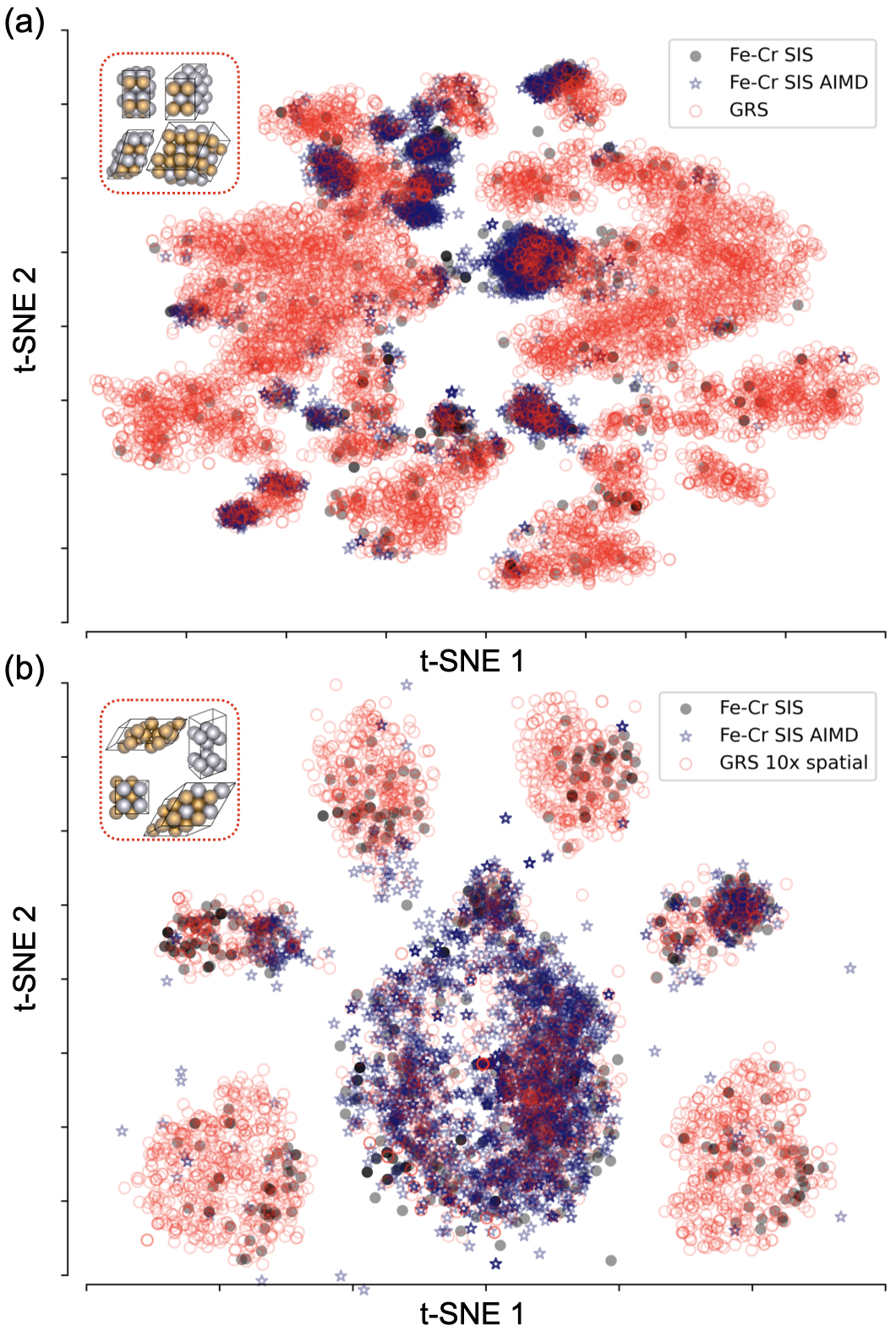}
    \caption{Comparison of two GRS sets, one with more chemical sampling (a) and the second more spatial sampling (b). This demonstrates the tuneable chemical versus spatial sampling of GRS methods and compares it to state-of-the-art fixed-lattice structure enumeration (black circles) and AIMD simulations (blue stars). }
    \label{fig:repro_multi}
\end{figure}

Two generalized representative structure sets generated with different degrees of spatial vs chemical sampling. For both sets, the sign of the inter and intra-structure loss are reversed to favor structures deviating from a random $Fe_{0.5}Cr_{0.5}$ BCC alloy target.  
The first set is was comprised of 1350 atomic structures (9482 atomic environments). The second set was generated with 600 GRS structures (3770 atomic environments).
By using negative pre-factors for both the inter-structure and intra-structure contributions to the GRS effective potential, structures may be enumerated that are \textit{dissimilar} to the random alloy target. However, to demonstrate the tuneable sampling for chemical versus spatial degrees of freedom, the spatial sampling probability for the second set was 10 times the spatial sampling probability for the first GRS set.
Finally a sampling of \~ 1200 AIMD (NVT 300 K) snapshots for SIS structures was also analyzed with the same ACE descriptor set for comparison.

Rather than comparing the high-dimension descriptor space for the fixed-lattice SIS, the AIMD, and the GRS, t-Distributed Stochastic Neighbor Embeddings (t-SNE) were performed for dimensionality reduction and clustering. This was done using the OpenTSNE package. 
In Fig. \ref{fig:repro_multi}(a), a common t-SNE embedding for the fixed-lattice SIS, the AIMD snapshots, and the first GRS set in black circles, open blue stars, and open red circles, respectively. The embedding for the most varied dataset amongst those three was used to compare all three datasets in the same reduced space. 
While this first GRS set spans many new regions of descriptor space that the SIS and AIMD trajectories do not, it doesn't span all of the thermally accessible states sampled with AIMD. 
By increasing the spatial sampling in the second GRS set, Fig. \ref{fig:repro_multi}(b), more of the thermally accessible states sampled in AIMD are spanned in the GRS set. This is accomplished while still sampling some new regions in descriptor space along with the regions spanned by the SIS. The insets in Fig. \ref{fig:repro_multi} show atomic structures for the GRS sets that highlight how the first samples more chemical degrees of freedom (with varied chemistries in structures resembling cubic crystal parent lattices) while the second samples more spatial degrees of freedom (with larger distortions from cubic parent lattices).

If spatial sampling is too low, the GRS may not sample as many thermally accessible states seen in AIMD. However, this is tuneable in GRS; more AIMD states may be recovered by increasing spatial sampling. In both examples, it can be seen that the GRS methods help sample new regions of descriptor space. One important result that the t-SNE embeddings highlights for each GRS set is that some of the fixed-lattice alloy structures and some of the AIMD snapshots are exactly reproduced. 
The GRS reproduces some crystal alloy and thermally accessed structures exactly, but also interpolates between them in chemical and spatial degrees of freedom. 

\bibliography{ace_sym.bib}
